\documentclass[conference]{IEEEconf}

\input epsf
\usepackage{cite}
\usepackage{amsmath,amssymb,amsfonts}
\usepackage{graphicx}
\usepackage{textcomp}
\usepackage[labelformat=simple]{subcaption}

\usepackage[hidelinks]{hyperref}
\usepackage{xcolor}
\usepackage{soul}
\usepackage{makecell}
\usepackage{algorithm}
\usepackage[noend]{algorithmic}

\usepackage{amsmath}
\usepackage{array}

\renewcommand\thesection{\arabic{section}} 

\renewcommand\thesubsectiondis{\thesection.\arabic{subsection}}

\renewcommand\thesubsubsectiondis{\thesubsectiondis.\arabic{subsubsection}}

\begin{document}

\title{\textbf{\Large Efficient patient-centric EMR sharing block tree\\}}

\author{Xiaohan Hu$^{1,*}$,  Jyoti Sahni$^{1,*}$,  Colin R. Simpson$^{1,2,*}$,  Normalia Samian$^{3}$,  Winston K.G. Seah$^{1,*}$\\
	\normalsize $^{1}$School of Engineering and Computer Science, Victoria University of Wellington, Wellington, New Zealand \\
	\normalsize $^{2}$Usher Institute, College of Medicine and Veterinary Medicine, The University of Edinburgh, Edinburgh, United Kingdom\\
	\normalsize $^{3}$Department Of Communication Technology And Network, Universiti Putra Malaysia, Selangor Darul Ehsan, Malaysia\\
	\normalsize xiaohan.hu@vuw.ac.nz, jyoti.sahni@vuw.ac.nz, colin.simpson@vuw.ac.nz, winston.seah@vuw.ac.nz, normalia@upm.edu.my\\
	\normalsize *corresponding author
}


\maketitle
\begin{abstract}
Flexible sharing of electronic medical records (EMRs) is an urgent need in healthcare, as fragmented storage creates EMR management complexity for both practitioners and patients. Blockchain has emerged as a promising solution to address the limitations of centralized EMR systems regarding interoperability, data ownership, and trust concerns. Whilst its healthcare implementation continues to face scalability challenges, particularly in uploading lag time as EMR volumes increase. In this paper, we describe the design of a novel blockchain-based data structure, MedBlockTree, which aims to solve the scalability issue in blockchain-based EMR systems, particularly low block throughput and patient awareness. MedBlockTree leverages a chameleon hash function to generate collision blocks for existing patients and expand a single chain into a growing block tree with $n$ branches that are capable of processing $n$ new blocks in a single consensus round. We also introduce the EnhancedPro consensus algorithm to manage multiple branches and maintain network consistency. Our comprehensive simulation evaluates performance across four dimensions: branch number, worker number, collision rate, and network latency. Comparative analysis against a traditional blockchain-based EMR system demonstrates outstanding throughput improvements across all dimensions, achieving processing speeds $\nu\cdot n$ times faster than conventional approaches.

\end{abstract}
\IEEEoverridecommandlockouts
\vspace{1.5ex}

%
\IEEEpeerreviewmaketitle

\section{Introduction}
\label{intro}
Electronic medical records (EMRs) have revolutionized healthcare delivery by enhancing productivity, quality, and accessibility of services \cite{sheikh2021health}. However, challenges persist in EMR management, particularly concerning data ownership, privacy, and interoperability. Current healthcare institutions usually maintain EMRs in their centralized databases, where healthcare providers retain control over patient data (although patients may be able to access their EMRs). Hence, the ownership, integrity, and sharing flexibility cannot be guaranteed~\cite{hermansen2022developing}. Moreover, fragmented storage across institutions further complicates cross-institutional patient care and impedes scientific health data research that requires patient consent. Additionally, EMR uploading latency within healthcare systems consistently triggers public dissatisfaction, especially for patients who need cross-institutional medical care \cite{astle2023we}. 
 
Numerous policies have been announced to regulate EMR management. For example, the Health Insurance Portability and Accountability Act (HIPAA) \cite{annas2003hipaa} and the EU General Data Protection Regulation (GDPR) \cite{goddard2017gdpr} have established standards for data privacy and patient access. The 21st Century Cures Act \cite{kesselheim2016breakthrough} further emphasizes interoperability and patient-centric data control. These policies highlight the need for technological solutions that meet regulatory requirements and operational efficiency in healthcare delivery.

Blockchain is considered the preferable option to advance current EMR systems due to its decentralization, traceability, and tamper-proof properties \cite{santos2021towards}. The fundamental concept is uploading the EMRs' relevant data into the blockchain to facilitate EMR sharing efficiency and usability. Since blockchain ledger is maintained by every system participant, EMRs can be easily shared, whilst protecting the records from unilateral tampering and data loss caused by single-point of failure~\cite{blockchainsurvey2023}. However, most blockchain-based solutions target financial industries. Its application in medical data management is still in the research phase, facing several open challenges: (1) scalability limitations (block throughput and decentralization); (2) regulatory compliance (security and ownership); (3) integration with legacy healthcare IT infrastructures (flexible sharing) \cite{attaran2022blockchain}.

In this paper, we propose MedBlockTree, a novel blockchain-based data structure designed to solve the aforementioned challenges in EMR systems. Our main contributions are summarized as follows:
\begin{enumerate}
    \item \textbf{MedBlockTree}: A new block tree structure that builds upon the blockchain concept and incorporates the chameleon hash function \cite{krawczyk1998chameleon, li2022efficient}. MedBlockTree significantly improves the on-chain data processing speed, blocks per second (BPS), by concurrently processing multiple EMR blocks in one consensus round. The reduction in uploading latency considerably improves system efficiency.

    \item \textbf{EnhancedPro consensus}: A new consensus protocol called EnhancedPro is introduced to process multiple winner elections in one consensus round and ensure network consistency. Workers can elect the blocks' proposers autonomously. The recognized randomness of every consensus round improves the fairness and unpredictability of winners' elections.
    
    \item \textbf{Patients' awareness}: Newly added EMRs are acknowledged by patients since a collision block's generation relies on the patient's secret key. This key is only held by patients, hence the patients' awareness is strongly ensured. There will not be disputes regarding the generation of EMRs.
\end{enumerate}

The remainder of this paper is structured as follows. Section~\ref{related} reviews the current research on EMR sharing and indicates the research gaps. The preliminaries are introduced in Section~\ref{preliminaries}, followed by the MedBlockTree model and EnhancedPro protocol in Section~\ref{MedBlockTree}. Section~\ref{analysis} reports the performance analysis based on evaluation results. The paper is concluded in Section~\ref{conclusion}, with a discussion of potential future research.

\section{Related works}
\label{related}
Blockchain adoption in healthcare has been widely studied. This section reviews prior work and highlights the existing gaps. 
\subsection{Blockchain-based EMR sharing}
The classical design of a blockchain-based EMR sharing system is directly uploading EMRs to blockchain, by leveraging blockchain's distributed and tamper-proof characteristics to facilitate secure sharing and protecting the overall records. For instance, Johari et al.\cite{BLOSOM} proposed a Proof of Work (PoW) based medical data sharing framework to enhance the security and traceability of EMRs. However, the PoW consensus is known for its high resource consumption. Besides, timely on-chain updating with large-sized data is a challenge, since the synchronization of new blocks among participants will be time-consuming, potentially resulting in a long uploading latency \cite{santiago2020accelerating}. 

An advanced design combines the existing off-chain database to decrease the EMR uploading latency. SPChain \cite{zou2021spchain} employed a hybrid storage approach where EMR metadata is stored on-chain while original records remain in hospital databases. Proof-of-Reputation (PoR) is utilized in SPChain to replace energy-intensive PoW, ensuring efficient consensus while incentivizing honest participation. Its cross-institutional EMR sharing is achieved through proxy re-encryption, ensuring privacy-preserving data access control. Li et al. \cite{li2023blockchain} uploaded the hashes of the EMRs on a blockchain to create an EMR address ledger, thereby reducing the latency associated with data uploads. The actual EMR data is encrypted and stored in a distributed database, accessible only via smart contracts, which allows patients to retain full control over their data. Similarly, Ma et al. \cite{ma2024integrating} integrated blockchain and InterPlanetary File System (IPFS) \cite{benet2014ipfs} to increase blockchain efficiency. EMRs' hash addresses and smart contracts are stored on the index chain to enhance retrieval, while the raw records are maintained in an IPFS constructed by the hospitals.

\subsection{Blockchain scalability in healthcare}
Blockchain ensures network consistency among the participants by avoiding bifurcations since the forks will easily incur record disputes and reduce system security, particularly in financial areas \cite{sakurai2023impact}. In healthcare, data uploading efficiency is critical, especially in emergencies where timely access to patient information is essential for making rapid, informed decisions. 

Examining the previously discussed designs, EMR-relevant data in blockchain-based healthcare systems must first enter a data pool to await processing, hence ensuring the chronological order of EMRs. However, as the institutions expand and patient numbers increase, the waiting queue in this data pool grows substantially. This expansion directly degrades system efficiency, creating processing delays that compromise the performance of time-critical medical applications. The linear processing constraint becomes especially problematic in blockchain applications. Hence, a design that maintains the core characteristics of blockchain while enhancing system efficiency is needed to offer a viable solution to the bottleneck.

Sharding and Directed Acyclic Graphs (DAGs) \cite{DAG1992} are two popular approaches to solve the low throughput issue in conventional blockchain \cite{rao2024scalability}. Sharding divides the blockchain network into multiple groups, each capable of processing transactions independently. This method improves throughput by enabling intra-shard parallelism. However, sharding introduces complex cross-shard communication protocols, which can lead to latency and consistency issues \cite{kale2024sharding}. Besides, independent ledgers among different shards will also introduce extra risk for global record updates, for instance, record sequential and cross-shard records combination, which makes it not suitable in healthcare. In DAG-based structures~\cite{Spectre, DAGUSENIX2020}, transactions are arranged in a graph where multiple blocks can be confirmed concurrently. This design enhances throughput by checking whether the calculated transaction has enough relevant former parent transactions. Nevertheless, DAG is suitable for the domains where transaction confirmation depends on cumulative network activity rather than deterministic consensus \cite{raikwar2024sok}. In healthcare, this will compromise the records' sequence and auditability.

Furthermore, both solutions are primarily tailored to financial applications, where eventual consistency and minor delays are tolerable. In contrast, EMR systems demand deterministic finality, strict sequentiality, and the ability to prioritize critical patient data in emergencies. These medical-specific requirements remain largely unaddressed by existing scalability solutions, indicating a critical gap in adapting blockchain for healthcare use cases. Table~\ref{tab:related} exhibits the comparison of the above works regarding the EMR sharing requirements.

\renewcommand{\arraystretch}{1.2}
\newcolumntype{C}[1]{>{\centering\arraybackslash}p{#1}}
\begin{table}[]\footnotesize
\centering
\caption{Comparison of blockchain relevant designs}
\footnotesize
\begin{tabular}{|l|C{0.7cm}C{0.6cm}C{0.85cm}C{1.1cm}C{1.15cm}|} \hline
Designs & BPS & Secure & Flexible & Ownership & Emergency \\ \hline
Blockchain & + & \checkmark & \checkmark & \checkmark & × \\
Plus off-chain & ++ & \checkmark & \checkmark & \checkmark & × \\
Sharding & +++ & \checkmark & × & - & × \\
DAG & ++++ & × & - & - & × \\
\textbf{MedBlockTree} & ++++ & \checkmark & \checkmark & \checkmark & \checkmark \\ \hline
\multicolumn{6}{|l|}{+ : Throughput, \checkmark : Supported, × : Unsupported, - : Not applicable}\\\hline
\end{tabular}
\label{tab:related}
\end{table}

\section{Preliminaries}
\label{preliminaries}
This section outlines the underlying algorithmic concepts deployed in MedBlockTree.
\subsection{Chameleon hash function}
Adi Shamir and Yael Tauman first defined the chameleon hash algorithm \cite{li2022efficient}. The proposed algorithm is a cryptographic primitive that enables the signers to create an undeniable digital signature for a specific target. With the secret key held by the signers, the signature can also be transformed into a signature for different content \cite{krawczyk1998chameleon}. The general outline of the chameleon hash function includes the following algorithms:
\begin{enumerate}
        \item\textit{Key generation}  \mbox{$(p_k, t_k) \leftarrow H_{Gen.}(1^k)$}. $k$ is the security parameter. $h_k$ is the public key and $t_k$ is the secret key.
        \item \textit{Hash value} \mbox{$(h,\zeta) \leftarrow Hash(h_k, m)$}. The standard cryptographic hash function maps the input to a fixed-size hash value and a verification string.
        \item\textit{Verification}\mbox{$(True, False)\!\leftarrow\! Hash_{V\!erif.}(h_k, m, (h,\zeta))$}. $Hash_{Verif.}$ resembles a verification function in which the recipients use the $h_k$ and $m$ to confirm the legality.
        \item \textit{Collision}\mbox{$(\zeta)'\!\leftarrow\! ColliHash(t_k, m', (m, h,\zeta))$}. The $t_k$ holder uses $t_k$ and $m'$ to generate the collision for which: \mbox{$H_{Verif.}(h_k, m, (h, \zeta))$ = $H_{Verif.}(h_k, m', (h, (\zeta)'))$}.
\end{enumerate}

\subsection{Verifiable Random Function}
A Verifiable Random Function (VRF) is a cryptographic construct that produces a pseudorandom output $y$ and an accompanying proof $\pi$ derived from a given input $x$ and a secret key \small$S\!K$\normalsize. $\pi$ allows any verifier, possessing the corresponding public key \small$P\!K$ \normalsize to confirm the correctness of $y$ without disclosing $x$ and $S\!K$ \cite{micali1999verifiable}. This property of verifiable randomness ensures that VRFs provide a reliable and secure mechanism for randomness verification, making them highly applicable in blockchain consensus protocols \cite{guo2022continuous}.

\section{MedBlockTree}
\label{MedBlockTree}
In this section, we present an overview of the MedBlockTree design, including its core idea, main components (blocks and branches), and the corresponding consensus protocol (EnhancedPro). To address the inefficiencies associated with uploading large-sized data within a single block, we align with the concepts presented by Ma et al. \cite{ma2024integrating} and Li et al.\cite{li2023blockchain}, and only record the main index (metadata) in the system. Hence focusing on improving the performance of on-chain data processing. 

Each participant (patients, practitioners, consensus workers) must register with the authority and acquire the key pair denoted by \small$(P\!K_n, S\!K_n)$\ \normalsize generated via the chameleon hash Key pair generation algorithm \mbox{{$H_{Gen.}(1^k)$}}. To clarify the main targets, we assume:
\begin{enumerate}
\item $SK_n$ is securely held by the owner.
    \item Metadata is generated based on the EMR regularities compliance, and is easy to understand. 
    \item Raw medical data is stored off-chain in a secure and encrypted manner. 
\end{enumerate}
We provide examples involving patients $Alice, Bob, Claire,$ and $Daisy$, along with representative EMR metadata keywords such as $cold, Flu, COVID$ to further illustrate how MedBlockTree manages EMRs.

\subsection{Core idea}
MedBlockTree targets the efficient processing of on-chain metadata updates. Its fundamental concept is, instead of eliminating the bifurcations, it generates recognized branches by implementing the chameleon hash function in the blockchain's underlying algorithm. Through expanding the one-chain structure into a growing tree, the throughput, blocks per round, is increased from one to $n$ ($n$ is the branch number) and contributes to improving the data processing efficiency. Since the chameleon hash collision can only be found by combining the secret key, which is only held by the holder, the blocks are generated based on awareness, preserving the system’s trust and consistency.

\subsection{MedBlockTree blocks}
In MedBlockTree, there are two types of blocks: (1) New blocks generated by the winners elected through consensus protocols; (2) Collision blocks created by clinic practitioners.

\subsubsection{New blocks}
Blockchain structure is a strictly arranged sequence in which the blocks are connected by the previous block's hash value. Since the hash calculation is a one-directional operation, data integrity, credibility, immutability, and traceability can be highly assured. In contrast to most blockchain models in which the default hash function is SHA256, MedBlockTree replaces it with an identity-based chameleon hash function~\cite{li2022efficient} to accommodate its unique design. 

Similar to SHA256, the chameleon hash algorithm generates a unique hash value, denoted as $Chamehash$, which is collision-resistant without access to a secret key \cite{krawczyk1998chameleon}. In addition, the algorithm produces corresponding chameleon check strings, which are essential for validating hash correctness and enabling controlled collision generation.

\begin{figure*}[t]
    \centering
    \begin{minipage}{0.25\textwidth}
        \centering\includegraphics[width=0.58\textwidth,trim=0mm 2mm 0mm 1mm]{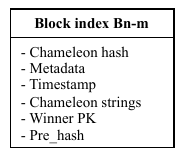}
        \caption{Block structure}
        \label{fig: MedBlockTreeblock}
    \end{minipage}\hspace{0.5ex}
    \hspace*{-5mm}
    \begin{minipage}{0.35\textwidth}
        \centering
        \hspace*{-15mm}
        \includegraphics[width=0.8\textwidth,trim=-0mm 0mm 2mm -3.5mm]{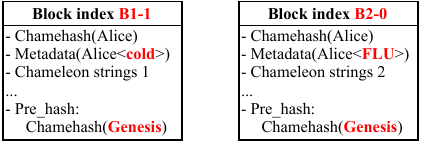}
        \caption{Alice's former and collision block}
        \label{fig: collision block}
    \end{minipage}
    \begin{minipage}{0.36419\textwidth}
        \centering
    \includegraphics[width=1.1\linewidth,trim=7mm -1mm -4mm -2mm]{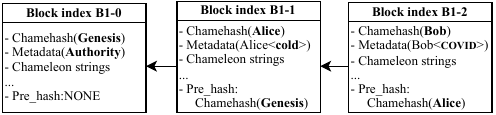}
    \caption{Example of the default chain \textit{B1}}
    \label{fig: example1}       
    \end{minipage}\hspace{1ex}
\end{figure*}

To distinguish blocks with identical hash values across branches, block indexes in MedBlockTree are designed as a tuple of the branch identifier $B(n)$ and the block sequence number $m$. Consequently, the block structure in MedBlockTree differs from traditional SHA256-based designs. Figure~\ref{fig: MedBlockTreeblock} illustrates the implemented block format.

\subsubsection{Collision blocks}
Normally, some patients require ongoing treatment, resulting in the generation of new medical records over time. However, when the metadata pool experiences a long waiting queue, newly created EMRs may suffer from significant upload delays due to the blockchain's sequential consensus process, which is unsuitable for urgent care or emergencies. 

MedBlockTree introduces the \textit{collision block} concept to support emergency suitable design. A collision block is a specially crafted block that shares the same hash value as a previous block but contains new metadata. As introduced in Section~\ref{related}, the chameleon hash algorithm enables controlled collision generation. Specifically, the creation of a collision block involves leveraging the chameleon check strings and content from the patient’s prior block, in combination with the patient’s private key ($SK$) and the new metadata. In EMR sharing, one of the primary concerns is data ownership \cite{hermansen2022developing}. We assume that if patients are aware of the EMRs awaiting upload, and the corresponding collision blocks are securely and legitimately generated by authorized practitioners, then these records can be considered validated. Consequently, if a returning patient's newly generated metadata is default validated, it can be prioritized and uploaded in the next consensus round. To illustrate this process, we use $Alice$ (first patient) and keywords $cold, Flu$ as an example to show the collision block generation.

\textit{\textbf{Example}}: Suppose Alice returned to the clinic at a later time. During the treatment, the doctor retrieved Alice's former medical records by \small$M\!eta(Alice\langle cold\rangle)$ \normalsize as a reference, and generated a new EMR with corresponding metadata pack \small$M\!eta(Alice\langle{Flu}\rangle)$\normalsize. Based on the contents in the former block \small$Alice\langle cold\rangle$ \normalsize and her secret key \small$S\!K_{Alice}$\normalsize, the system can easily find a collision (same hash value different chameleon strings) for the new metadata \small$M\!eta(Alice\langle \text{Flu}\rangle)$\normalsize, where:

\vspace{-10pt}
\small\begin{align*}
&\text{Collision}(\zeta_2) \leftarrow 
\text{ColliHash}(S\!K_{\text{Alice}},\ M\!eta(\text{Alice}\langle \text{Flu} \rangle),\notag \\
& M\!eta(\text{Alice}\langle \text{cold} \rangle),\ P\!K_{\text{Alice}},\ \zeta_1)
\end{align*}\normalsize

All the participants can validate $Collision(\zeta_2)$ by using the verification function $Hash_{Verif}$ to confirm:
\small\begin{align*}
&\text{Hash}_{\text{Verif.}}(P\!K_{\text{Alice}},\ M\!eta(\text{Alice}\langle \text{cold} \rangle),\ (\text{Chamehash}(\text{Alice}),\ \zeta_1)) = \notag \\
&\text{Hash}_{\text{Verif.}}(P\!K_{\text{Alice}},\ M\!eta(\text{Alice}\langle \text{Flu} \rangle),\ (\text{Chamehash}(\text{Alice}),\ \zeta_2))
\end{align*}
\normalsize
Since the collision block $Alice\langle Flu\rangle$ owns the same hash value as $Alice\langle cold\rangle$ has, and the new block is seen as validated, the pre-hash in block $Alice\langle Flu\rangle$ will also be the hash value in the genesis block (same as block $Alice\langle cold\rangle$). The newly generated block and $Alice$'s former block are shown in Figure~\ref{fig: collision block}.

\subsection{MedBlockTree branches}
MedBlockTree is constructed by the default chain \textit{B1} and the generated branch $B(n)$, $n>1$.  
\subsubsection{Default chain - B1}
The default chain generation is similar to the other EMR blockchain sharing systems. In an authorized hospital, patient $p_i$ provides \small$P\!K_{p_i}$\normalsize to doctors. After treatment, doctors generate medical records $Raw_{p_i}$ of $p_i$ and the corresponding metadata $M\!eta_{p_i}$. The treatment system packs \small$M\!eta_{p_i}$\normalsize with \small$P\!K{p_i}$\normalsize, doctor $PK$ and the timestamp as a metadata pack, and these assembled packs will be dispatched to the waiting queue, within the metadata pool awaiting to append to the chain. The genesis block is built by the authority as the starting point. And the following blocks are strictly connected by the previous block's chameleon hash, as shown by the following example.

\textit{\textbf{Example}}: In the default phase, we assume the sequence in the metadata pool is: \small$M\!eta(Alice\langle cold\rangle)$, $M\!eta(Bob\langle COVID\rangle)$\normalsize. The participants in MedBlockTree join the consensus competition to win the opportunity to be the block proposer. After two consensus rounds, the default chain \textit{B1} is extended to three blocks, as shown in Figure~\ref{fig: example1}.

\subsubsection{Branch - B(n)}
\label{branch2}
\begin{figure}[b]
    \centering
    \includegraphics[width=.85\linewidth]{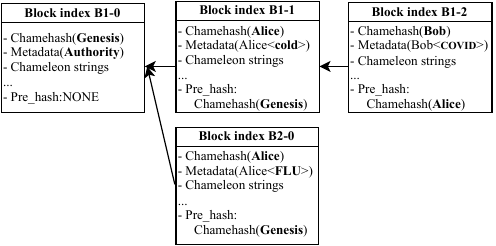}
    \caption{Example of MedBlockTree with two branches}
    \label{fig:B2}
\end{figure}

Since patients' collision blocks share the same hash and pre-hash values as their corresponding former blocks, they are positioned at the same level in the chain. This structural overlap forms what is traditionally referred to as a fork in blockchain systems. However, in MedBlockTree, patients are aware of the creation of collision blocks, and those blocks are default as validated. To further enhance the efficiency of metadata processing, those default validated collision blocks are designated as the origin points for new branch growth. Following the given examples of $Alice'$ blocks and default chain, a new branch named \textit{B2} now has been added to the default chain \textit{B1}, as shown in Figure~\ref{fig:B2}. By analogy, every time an existing patient returns, a new branch will be generated, and a blockchain-based structure will quickly grow into a tree, namely, MedBlockTree.

\subsection{EnhancedPro protocol}
The consensus protocol is a recognized algorithm that every consensus competition participant needs to comply with to ensure consistency \cite{xu2023survey}. Various consensus protocols have been proposed for blockchain systems, yet one of the primary objectives is to prevent forks. Therefore, they are not well-suited for the requirements of MedBlockTree, which embraces branching structures.

To match the parallel processing of multiple blocks in one consensus round and maintain MedBlockTree network consistency, we build upon an advanced consensus protocol, viz., Enhanced Tendermint \cite{bui2023enhanced},  by refining its underlying algorithm, and designing a multiple winners consensus protocol - EnhancedPro. Similar to Enhanced Tendermint, EnhancedPro is also a hybrid of Proof of Stake (PoS) and Byzantine Fault Tolerant (BFT). Consensus workers must undergo a bootstrapping phase (qualification review and deposit tokens) to join the consensus process. EnhancedPro operates in three key stages: (1) PoS-based self-processing, (2) BFT-based network consensus, and (3) Tree update.

\subsubsection{Self-processing}
Self-processing refers to the period during which each participant autonomously collects round-specific information to conduct winner-election, self-checking, and generate the proposals (only for the winners). 

PoS is a consensus mechanism in which workers are selected to propose blocks based on the amount of stake they provide, cryptocurrency or reputation, for instance, promoting energy efficiency without using the computing mining in PoW. However, a critical drawback of PoS is its tendency toward centralization, as nodes with higher token balances have a greater chance of being selected as winners \cite{xu2023survey}.

To improve the unpredictability and fairness of winner elections, we introduced sub-nodes and randomness. It is worth emphasizing that in MedBlockTree, the core mechanism for parallel processing of multiple blocks in one consensus round is that each branch can be seen as a block append point. Hence, a recognized algorithm for estimating winners for each branch also needs to be considered.

\textbf{Sub-nodes}: Given MetaPool $\{meta_0, meta_1, $\ldots$, meta_\iota\}$, workers' public keys \small$\{P\!K_{W_0}, P\!K_{W_1}, $\ldots$, P\!K_{W_k}\}$\normalsize, deposit tokens \small\{$\tau(W0)$, $\tau(W1)$, $\ldots$, $\tau(W_K)$\}\normalsize, and total deposits \small$\sum(\tau_{m})$\normalsize, the new recognized worker list \small\{$W_0$, $W_1$, $\ldots$, $W_K$\normalsize\} is generated through the default sorted configuration in the MedBlockTree system. The sub-nodes and sub-nodes' indexes are defined as follows: sub-nodes \{$\omega_0$, $\omega_1$, $\ldots$, $\omega_k$\}, with 
sub-nodes' indexes \{0, 1, $\ldots$, k\}. For example, $W_1$ deposit two tokens and sub-nodes of $W_1$ are $\omega_0$ and $\omega_1$, with indexes 0 and 1.

\textbf{Randomness}: Each worker maintains a former blocks map \{\textit{B1}: $block_x$, \textit{B2}: $block_y$, $\ldots$, \textit{B(n)}: $block_z$\} to record the latest blocks for branches. The round randomness is determined by two public parameters: $\zeta$ and $n$. For which $\zeta$ is the chameleon check string in the former block, and $n$ is the corresponding branch number.

\textbf{Winner election}: For each branch $B(n)$, workers will perform a module calculation. The result corresponds to the index of the chosen sub-node, which maps to the worker $W_k$. This $W_k$ will be the block proposer of $B(n)$. As shown in Equation~\eqref{eq:wi}.
\begin{equation}
\label{eq:wi}
     \omega_i = Hash(\zeta, n) \mod \sum(\tau_{m})
\end{equation}

Furthermore, to prevent malicious workers from broadcasting proposals and compromising network consistency, the VRF is deployed to generate the provable random values, $V\!RF_y$ and $V\!RF_\pi$, for winners. The other workers can validate the block by the winners' $P\!K$s and the proofs. The overall algorithm of self-processing is shown in Algorithm~\ref{alg:winner}.

{\renewcommand{\baselinestretch}{1.1}
\renewcommand{\algorithmicrequire}{\textbf{Input:}}
\renewcommand{\algorithmicensure}{\textbf{Output:}}
\begin{algorithm}[]
    \footnotesize
    \caption{MedBlockTree Self-processing Algorithm}
    \label{alg:winner}
    \begin{algorithmic}[1] 
        \REQUIRE former blocks map, sub-node indexes, workers' list
        \ENSURE  winners map=\{\textit{B1}: $winner_1$, \textit{B2}: $winner_2$, ..., \textit{B(n)}: $winner_n$\},\\ block proposals, \(V$\!$RF_y\text{s}, V$\!$RF_\pi\text{s}\)
        \STATE \textbf{New consensus round start:}
        \STATE \textbf{Step 1: Winners election}
        \FOR{each branch $B(n)$}
            \STATE Randomness of \(B(n)\) = \(\langle (B(n):block_i:\zeta), n \rangle \)
            \STATE winner's sub-node index = Randomness of $B(n) \mod \sum(\tau_{m})$
            \ENDFOR
        
        \STATE \textbf{Step 2: Self-checking}
        \FOR{each branch $B(n)$}
            \STATE current index (CurInd) = 0
            \FOR{each worker \(W_i\)}
                \IF{winner's sub-node index\:$ \in[\text{CurInd}, \text{CurInd} + \tau(W_i))$}
                    \STATE winner for \( B(n) = W_i\)
                    \STATE Append map $\langle B(n): W_i \rangle$ to winners map
                    \STATE \textbf{break}
                \ENDIF
                \STATE current index += $\tau(W_i)$
            \ENDFOR
        \ENDFOR
        \STATE \textbf{Step 3: New proposals}
        \FOR{each branch $B(n)$ with meta information $meta_\iota$}
            \IF{$n = \iota$}
                \STATE $W_i \gets$ winners map[$B(n)$]
                \STATE block of \( B(n) \) = \( \text{createBlock}(winner: W_i, meta_\iota) \)
                \STATE \( V\!RF_y \gets \text{VRFProve}(S\!K(W_i), \text{Randomness of } B(n)) \) 
                \STATE \( V\!RF_\pi \gets \text{VRFProve}(S\!K(W_i), V\!RF_y, \text{Randomness of } B(n)) \) 
            \ENDIF
        \ENDFOR
        \RETURN winners map, block proposals, \(V\!RF_y\text{s}, V\!RF_\pi\text{s}\)
    \end{algorithmic}
\end{algorithm}}
    
\subsubsection{Network consensus}
We adopt the BFT protocol to enhance efficiency and maintain network consistency. It contains three steps: (1) Winners broadcast the proposals and VRF verification strings they generated through self-processing; (2) Workers validate the proposals through the winner's $P\!K$ and VRF strings by the verification algorithm in Equation~\eqref{vrfverify}, and broadcast pre-vote; (3) Commit voting and the finalization of new blocks. Note that the ultimate $True$ of the blocks validation and finalization depends on the voting power of the votes. The worker's voting power equals the tokens deposited, and the requirement is at least 2/3 of \small$\sum(\tau_{m})$ \normalsize agree with the proposal. 

\small
\begin{equation}
\label{vrfverify}
        \text{Verify}(P\!K(W_i), V\!RF_y, \zeta, n, V\!RF_\pi) \rightarrow \text{True or False}
\end{equation}
\vspace{-5pt}

\begin{figure}
    \centering
    \includegraphics[width=.8\linewidth]{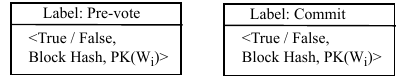}
    \caption{Pre-vote and Commit vote}
    \label{fig: votes}
\end{figure}

\begin{figure}
    \centering
    \includegraphics[width=0.85\linewidth]{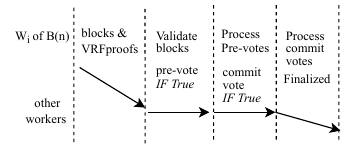}
    \caption{Events in networks}
    \label{fig: network}
\end{figure}

The pre-vote and commit vote structures are shown in Figure~\ref{fig: votes}. New blocks map BlockMap = \small $\{(\textit{B1}: \text{Block}_1,\; \textit{B2}: \text{Block}_2,\; \ldots,\; \textit{B}(n): \text{Block}_n)\}$
\normalsize will eventually be committed after the network events shown in Figure~\ref{fig: network}.

\subsubsection{Tree update}
Tree updating contains two parts: new blocks appending and collision blocks claiming. The new blocks will be appended to the MedBlockTree according to the BlockMap. The collision blocks created by the practitioners are based on the contents of the patients' previous blocks, as aforementioned. They are initially validated and then uploaded to the data pool to await processing. To support the growth of MedBlockTree, the collisions are updated at the end of each consensus round. Analogously, whenever an existing patient returns, a new branch is generated, enabling the blockchain structure to evolve rapidly into a tree, hence forming the MedBlockTree. This example continues from Example~\ref{fig:B2}, and is provided to illustrate this phase:

\textbf{Example}: Suppose there is a collision block of Bob \small$Colli(Bob\langle Flu\rangle)$\normalsize. Daisy and Ella are two new patients with meta packs \small$M\!eta(Claire\langle Flu\rangle)$ \normalsize and \small$M\!eta(Daisy\langle COVID\rangle)$\normalsize. In a new consensus round, workers of MedBlockTree propose the blocks of Claire and Daisy based on the EnhancedPro consensus protocol and get the final BlockMap = \small$\{$\textit{B1}$: \!Block(Claire\langle Flu\rangle),~$\textit{B2}$: \!Block(Daisy\langle COVID\rangle)\}$\normalsize. 
\begin{figure}[b]
    \centering
    \includegraphics[width=.9\linewidth]{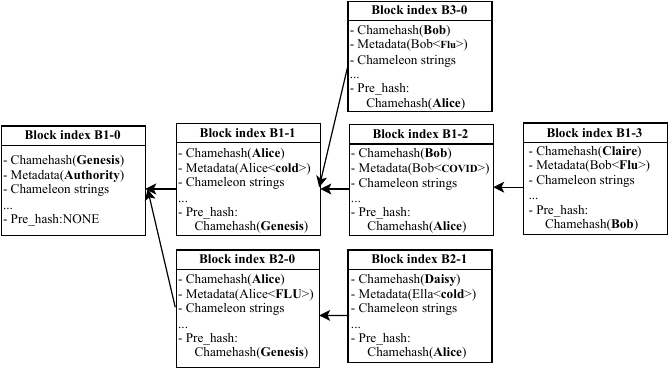}
    \caption{Example of updated MedBlockTree}
    \label{fig: MedBlockTree tree}
\end{figure}
Following the MedBlockTree we got in Section \ref{branch2} with two branches: \textit{B1} and \textit{B2}, Block $Claire\langle Flu\rangle$ will be added after Block $Bob\langle COVID\rangle$, while Block \small$Daisy\langle COVID\rangle$ \normalsize will be appended after Block $Alice\langle Flu\rangle$. As for \small$Colli(Bob\langle Flu\rangle)$\normalsize, with the same pre-hash of Block \small$Bob\langle COVID\rangle$\normalsize, it will be added as the starter of a new branch $B3$. The graphical description of the MedBlockTree is exhibited in Figure~\ref{fig: MedBlockTree tree}. 

\section{Performance analysis}
\label{analysis}
In this section, we evaluate the scalability of MedBlockTree by comparing its performance with a blockchain system implementing the original Enhanced Tendermint protocol across five key dimensions.

We deploy our nodes on virtual machines hosted by Catalyst Cloud \cite{catalystcloud}, organized into three clusters: patient, doctor, and worker. The worker nodes use sockets to conduct peer-to-peer communication, with network delays configured to simulate real-world conditions. RocksDB \cite{rocksdb} is deployed as the worker's database. The doctor and patient clusters are responsible for generating metadata and collision blocks, which are subsequently announced to the worker cluster for processing. All experiments were conducted under identical configurations and network conditions to ensure fair comparison. The simulation configurations are listed in Table \ref{tab:simu}.
\begin{table}[]\small
\renewcommand{\arraystretch}{1.15}
    \centering
    \caption{Simulation environments}
    \begin{tabular}{|p{2.5cm}|p{4.5cm}|}\hline
    Environment & Description \\\hline
    Platform & Catalyst Cloud \\
    Operating System & ubuntu (24.04 64), 2 core, 4 RAM\\
    Network status & 100ms / 200ms latency \\
    Data set & 2000 generated EMRs \\
    Database & RocksDB, Redis\\ \hline
    \end{tabular}
    \label{tab:simu}
\end{table}

\subsection{Benchmark test}
We first conduct a benchmark test (10000 rounds) to evaluate the time consumption for block formation, generation and upload in the Enhanced Tendermint blockchain using SHA256 \footnote{Secure Hash Algorithm 256-bit, which is widely used in blockchain-based systems, such as Bitcoin, Tindermint}, and in MedBlockTree employing an identity-based chameleon hash \cite{li2022efficient}. The results are summarized in Table~\ref{tab:bench}. SHA256, Chame. and Colli. indicate the time consumption on formatting a SHA256 block, a chameleon hash block, and a collision block respectively. The chameleon hash block generation consumes a bit longer due to its higher computational complexity. Yet we can see with the required information that, collision block generation time indicates a relatively low delay. BC (0.1 s) and BC (0.2 s) depict the average round time for the blockchain-based system with a network latency of 100 ms and 200 ms. DB denotes the average time needed to upload a block into RocksDB. 
\renewcommand{\arraystretch}{1.15}
\begin{table}[b]\small
    \centering
        \caption{Benchmark test (in seconds)}
    \begin{tabular}{|p{1cm}|c|c|c|c|c|}\hline
        SHA256 & Chame. & Colli. & BC (0.1) & BC (0.2)& DB\\\hline
        0.055 & 0.062 & 0.027 & 0.370 &0.676 & 0.011\\\hline 
    \end{tabular}
    \label{tab:bench}
\end{table}

\subsection{D1: Branch numbers}
Throughput is a main index to evaluate the system's performance. In this section, we simulate the MedBlockTree with various branch numbers and compare the efficiency, measured in blocks per second (BPS), against that of the Enhanced Tendermint blockchain. Four workers participate in the consensus competition with a 100 ms network latency.

\begin{figure}[]
    \centering
        \includegraphics[width=\linewidth, trim=0pt 0pt -0pt 0]{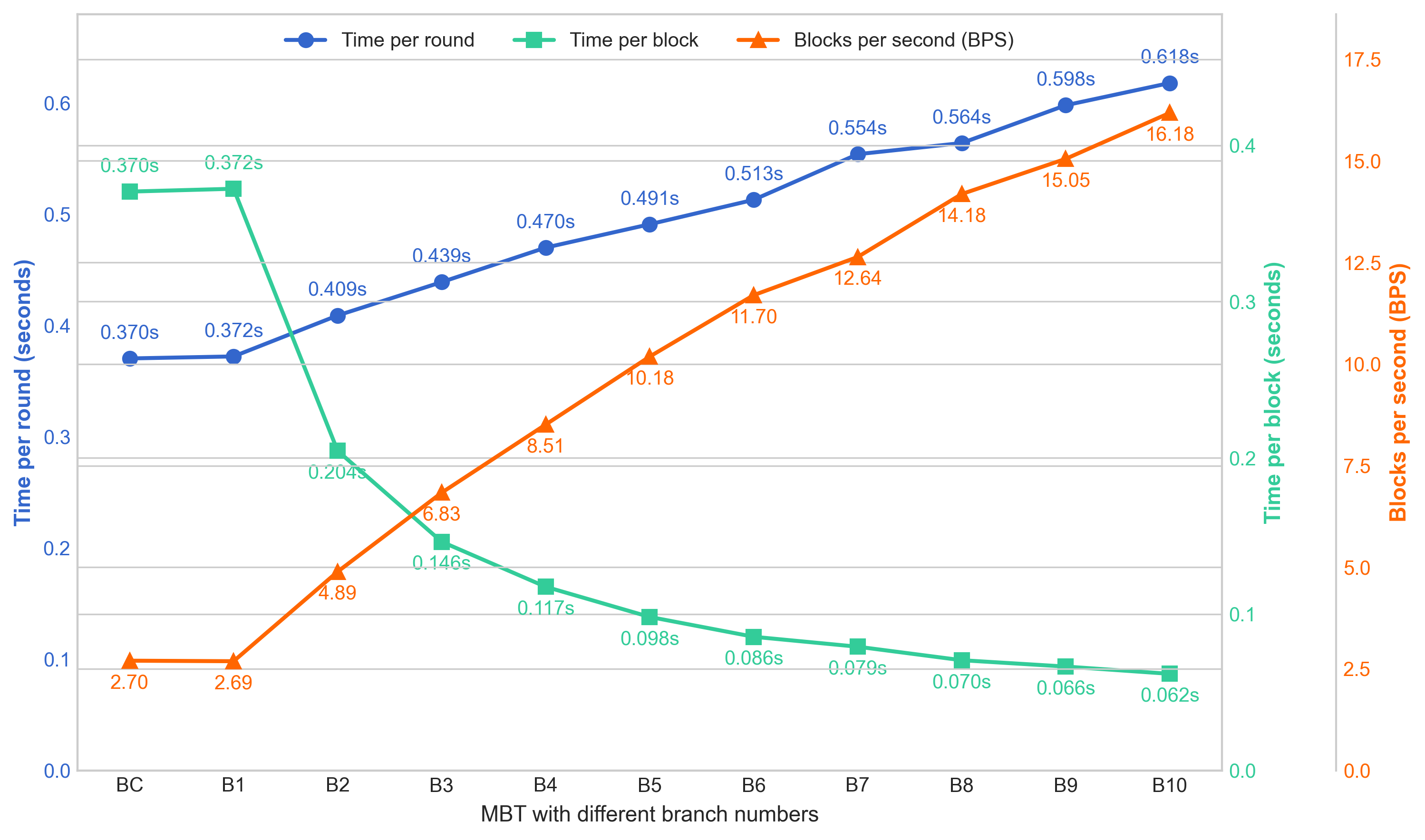}
    \caption{MedBlockTree performance with increasing branch numbers (BC: Enhanced Tendermint blockchain; B1-B10: MedBlockTree with different branch numbers.)}
    \label{fig:mainfig}
\end{figure}

Figure~\ref{fig:mainfig} exhibits the results of a blockchain (BC) and a MedBlockTree with different branch numbers (B1-B10) through three metrics: Time per consensus round (blue), Time per block (green), and Blocks per second (orange). As the number of branches increases, the consensus round time shows a slight increase, reflecting the additional blocks processed in each round in MedBlockTree. However, processing efficiency improves significantly: Time per Block decreases sharply from 0.372 s to 0.062 s, while BC is constantly 0.370 per round. Consequently, the system's throughput exhibits substantial enhancement, with BPS rising from 2.70 in blockchain to 16.18 under the 10-branch configuration (B10) in MedBlockTree.

Hence, with the branch number increasing, MedBlockTree can significantly improve the system efficiency, particularly on BPS, with a slight increase in consensus round time.

\subsection{D2: Network latency}
\begin{figure}[]
    \centering
    \includegraphics[width=\linewidth, trim=0pt 0pt 0pt 0]{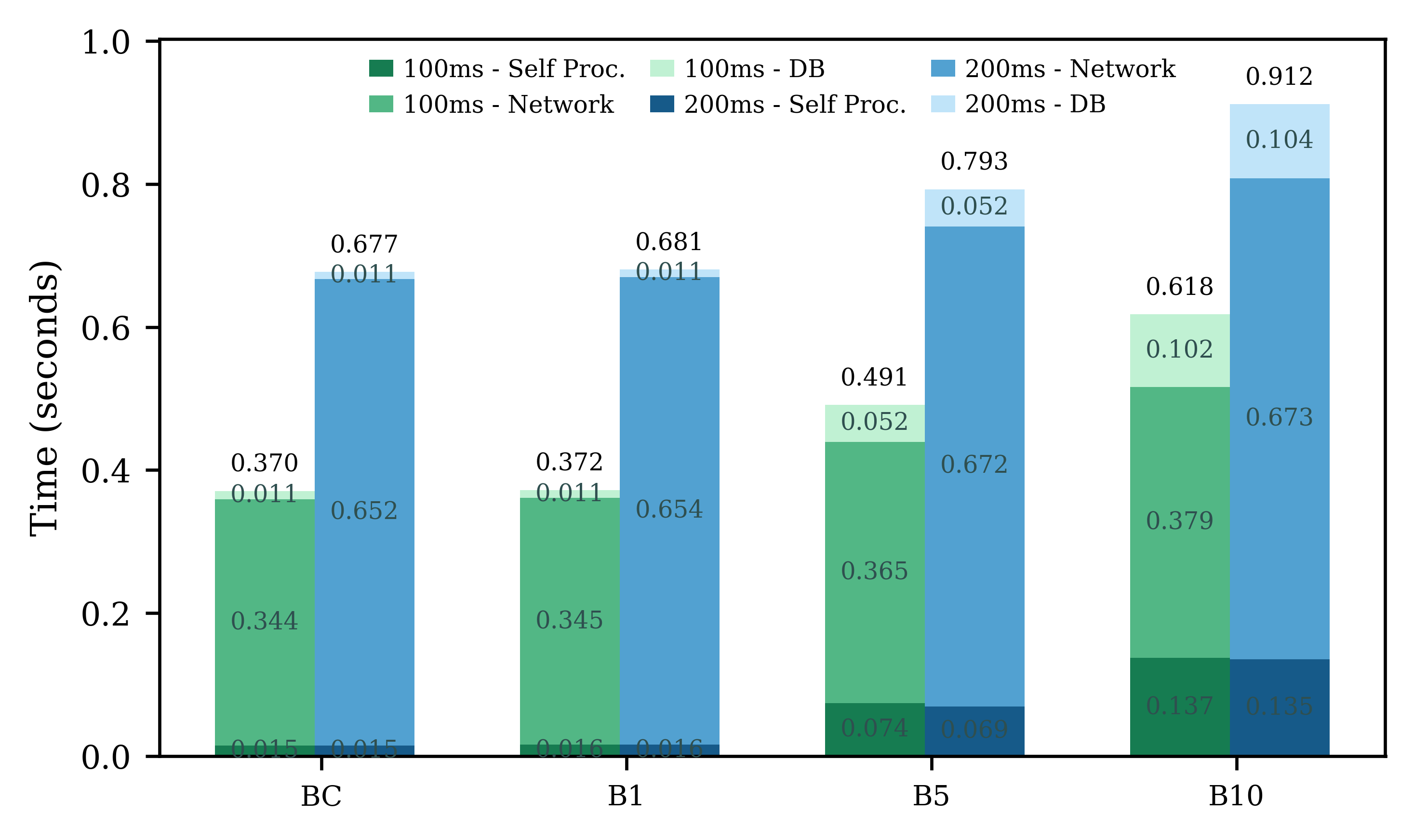}
    \caption{Durations comparison under different network status}
    \label{fig: duration}
\end{figure}

To clearly compare how network conditions affect performance, we conduct a duration analysis that includes self-processing, network consensus, and database processing times for each design. The results are shown in Figure~\ref{fig: duration}, where the green and blue columns represent network latencies of 100 ms and 200 ms, respectively. 

Overall, network latency influences the efficiency of both designs. In MedBlockTree, as the number of branches increases, all three durations exhibit a slight increase in time consumption due to the larger number of tasks processed per round. However, compared to the blockchain design, MedBlockTree leverages the consensus process more effectively by handling multiple blocks per round, thereby achieving significantly higher efficiency.

Additionally, in our implementation, RocksDB appends blocks sequentially, resulting in a linear increase in database update time. If this duration is disregarded (as it falls outside the scope of our research), MedBlockTree demonstrates even greater performance benefits. Notably, the epoch durations of the blockchain (BC) and MedBlockTree with a single default branch are nearly identical, with only a minor difference attributable to the larger block size in MedBlockTree\footnote{The block size is larger than that of SHA256-based blocks primarily due to the verification strings generated by the chameleon hash algorithm, which are essential for collision detection.}.

\subsection{D3: Number size}
In this section, we evaluate the scalability of MedBlockTree across varying network sizes (4, 8, 12, and 16 nodes). During the tests, the doctor and patient clusters generate collision blocks at a fixed rate of one every 60 seconds. The blockchain system with 4 nodes is used as the baseline to provide a clear performance comparison. A total of 2000 metadata entries are processed during each test.

\begin{figure}[]
    \centering
    \includegraphics[width=.95\linewidth, trim=0mm 0mm 0mm 0mm]{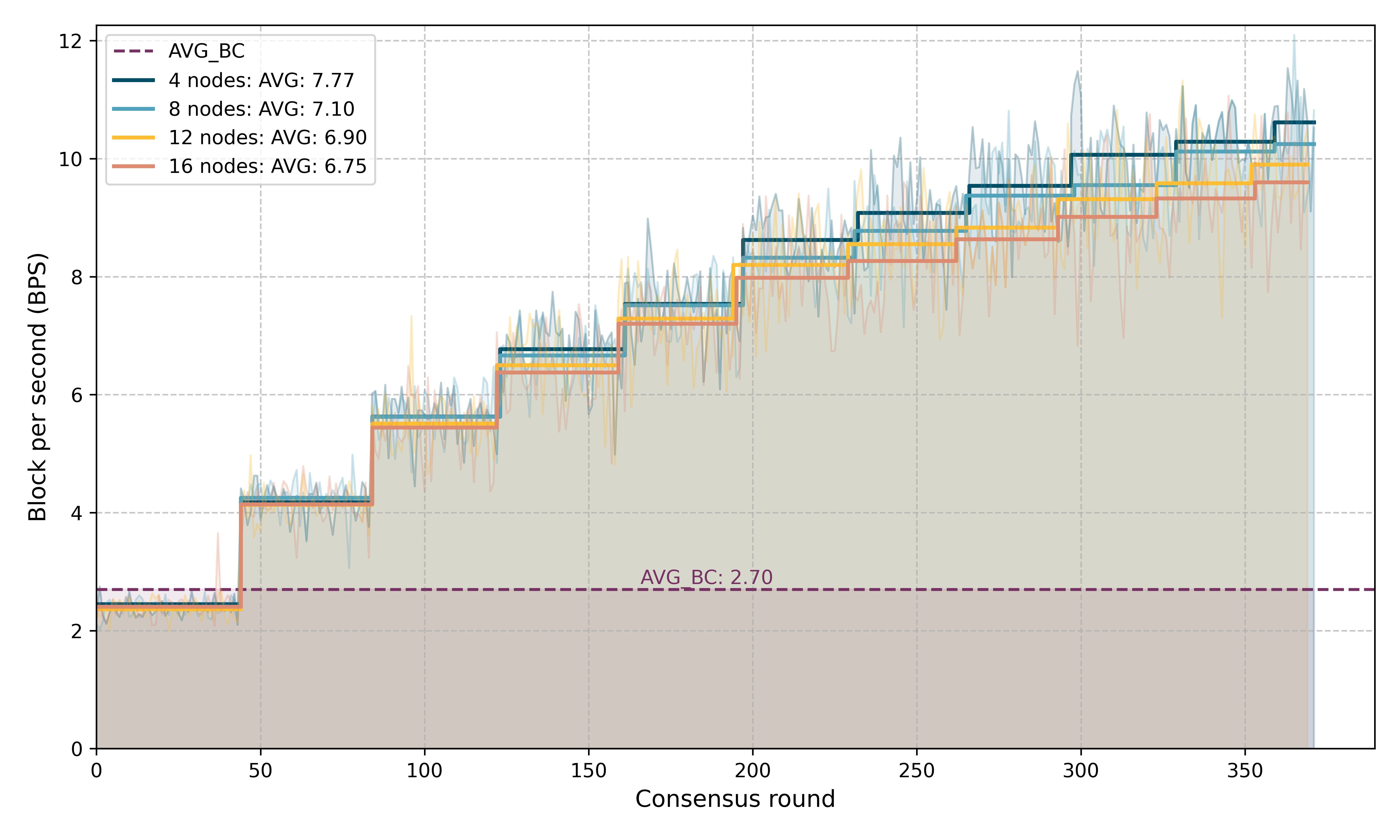}
    \caption{Performance comparison with various network sizes}
    \label{fig: nodes}
\end{figure}

In Figure~\ref{fig: nodes}, we exhibit the results by showing the BPS trend under different network sizes. It is obvious that the BPS across all network sizes is increasing regularly with the generation of new branches, which is outstanding compared to a constant processing speed in blockchain. However, the BPS of a larger network size MedBlockTree is smaller than that of one with fewer participants. For instance, the average BPS of MedBlockTree with 4 nodes is 7.77, whereas with 16 nodes it drops to 6.75. This is attributed to the increased communication overhead required for network consensus as more participants are involved, which is also a normal phenomenon in the blockchain systems \cite{rao2024scalability}. Nevertheless, even with 16 nodes, MedBlockTree significantly outperforms the baseline blockchain with only 4 nodes, with 2.5 times higher BPS.
We also conclude the detailed comparison results overall processing time (Overall), average BPS (AVGBPS), average time per block (AVGBP), and number of rounds (Rounds) in Table~\ref{tab:nodes}. In which with the same network size of 4 nodes (4n), the MedBlockTree (MBT) performs 2.87 times efficiently in processing all metadata, even though MedBlockTree has 16 nodes, it still runs 2.13 times faster. Hence, with more participants in the network, the performance may decrease a bit due to more network communication, but still indicates a great improvement compared to one chain structure.
\newcolumntype{C}[1]{>{\centering\arraybackslash}p{#1}}
\begin{table}[]
    \centering
        \caption{Performance details with various network sizes}
    \begin{tabular}{|l|C{1.2cm}C{1.2cm}C{1.2cm}C{1.2cm}|}\hline
    Syetem & Overall & AVGBPS & AVGPB & Rounds\\\hline
        BC (4n) & 738.01s & 2.70 & 0.37s &2000 \\
        MBT(4n) & 257.43s& 7.77 & 0.13s & 371\\
        MBT(8n) & 276.63s & 7.23 & 0.14s & 371\\
        MBT(12n) & 307.22s & 6.51 & 0.15s &371\\
        MBT(16n) & 346.62s& 5.77 & 0.17s & 371\\\hline
    \end{tabular}
    \label{tab:nodes}
\end{table}

\subsection{D4: Collision rates}
The collision rate reflects how frequently collision blocks are generated, which corresponds to how often existing patients return for subsequent treatments in the context of MedBlockTree. This rate directly impacts the expansion speed of the MedBlockTree, as higher collision rates lead to more frequent branch creation, enabling more blocks to be processed in future consensus rounds. In this section, we simulate the system under four different collision rates (30, 60, 90, and 120 seconds) with 4 nodes and 100 ms network latency, to visually examine how this parameter impacts system performance.
\begin{figure}[]
    \centering \includegraphics[width=\linewidth,trim=0mm 0mm 0mm 8mm]{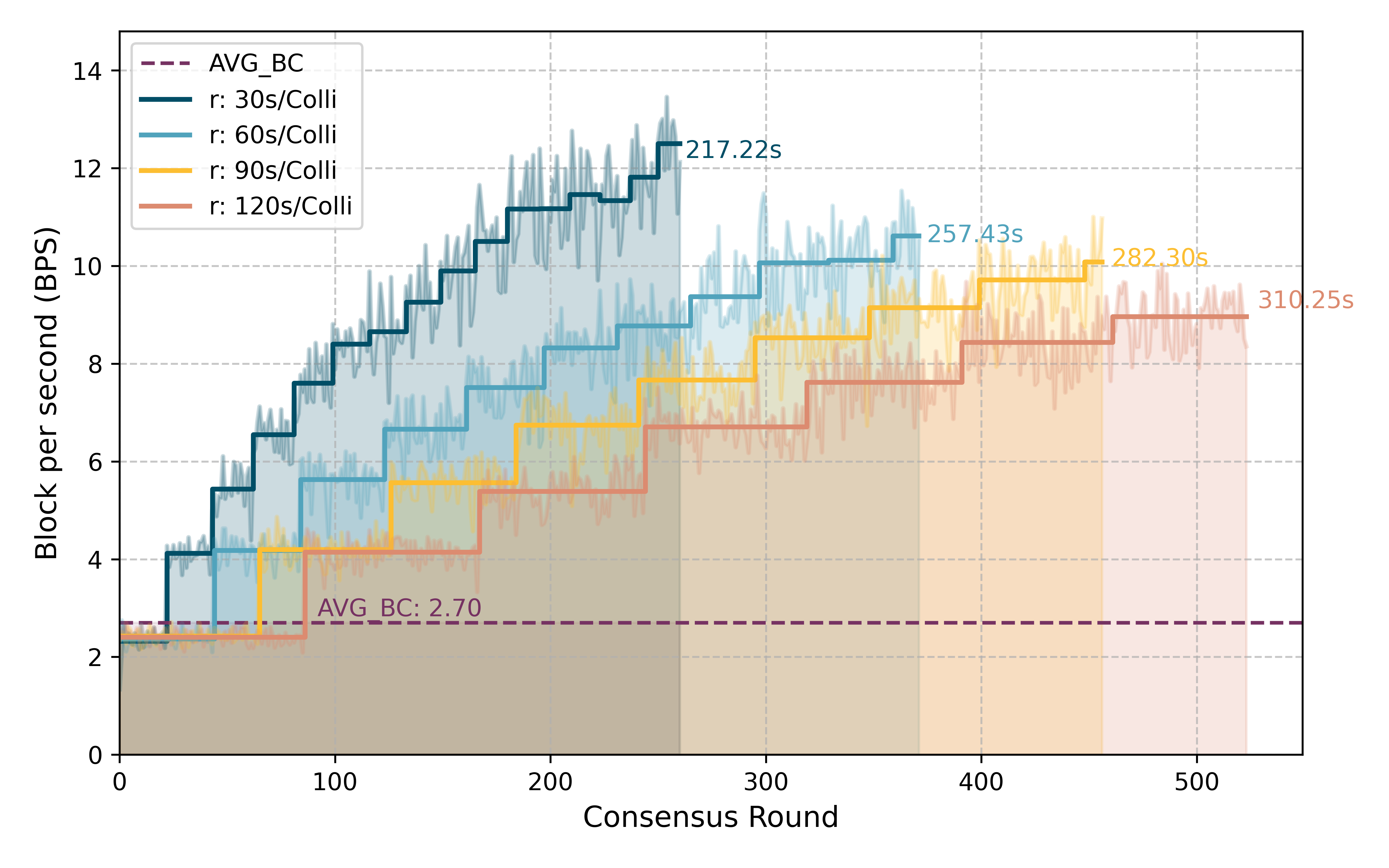}
    \caption{Performance comparison with various collision rates}
    \label{fig:rate}
\end{figure}
As shown in Figure~\ref{fig:rate}, the BPS trends for all four rates exhibit a similar stair-like growth pattern. The more frequently the collision is released, the steeper the steps become, leading to a more efficient processing of the overall dataset. When comes to a blockchain system, the consistent processing speed will largely influence the system's efficiency, which needs 738.01 seconds at 2000 rounds to finish. The detailed comparison is summarized in Table~\ref{tab:rate}. In a blockchain-based EMR system, the more subsequent treatments a patient requires, the longer the EMR uploading queue tends to be. In contrast, MedBlockTree leverages these frequent updates to improve system efficiency—more collisions lead to faster processing speeds, thereby further reducing lag time.

\newcolumntype{C}[1]{>{\centering\arraybackslash}p{#1}}
\begin{table}[]
    \centering
    \caption{Performance details with various collision rates}
    \begin{tabular}{|l|C{1.2cm}C{1.2cm}C{1.2cm}C{1.2cm}|}\hline
    Rate & Overall & AVGBPS & AVGPB & Rounds\\\hline
        BC (4n) & 738.01s & 2.70 & 0.37s &2000 \\
        MBT(30s) & 217.22s& 8.39 & 0.11s & 260\\
        MBT(60s) & 257.43s& 7.77 & 0.13s & 371\\
        MBT(90s) & 282.30s & 7.08 & 0.15s &456\\
        MBT(120s) & 310.25s& 6.45 & 0.17s & 523\\\hline
    \end{tabular}
    \label{tab:rate}
\end{table}

\subsection{D5: Fairness}
Fairness and unpredictability among workers are a vital concern of participants and a key contribution of Enhanced Tendermint. Our protocol, EnhancedPro is built upon Enhanced Tendermint. In this section, we measure the fairness among 4 nodes in both systems through the proportion each node is selected as the winner.

\begin{figure}[]
    \centering
    \includegraphics[width=0.95\linewidth, trim=0mm 5mm 0mm 0mm]{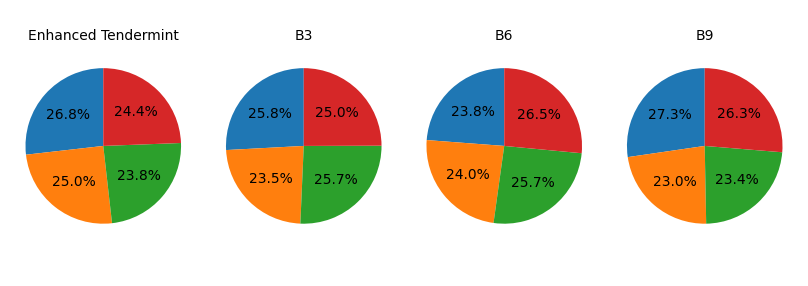}
    \caption{Fairness among workers}
    \label{fig: fairness}
\end{figure}

In EnhancedPro, the winner election for each branch is based on the chameleon string of the latest block, only till the beginning of the current round can the nodes predict the winners, decreasing the possibility of nodes attacking. Additionally, the chameleon string is generated through the hash function, ensuring the value's randomness. Hence, the EnhancedPro preserves the main characteristic from Enhanced Tendermint with multiple winners being elected in one round. Figure~\ref{fig: fairness} shows the comparison among Enhanced Tendermint and MedBlockTrees using examples of $B(3)$, $B(6)$, $B(9)$ (with fairness remaining consistent across other different branch numbers). Each colour in the pie chart represents one worker.
\section{Conclusion}
\label{conclusion}
In this paper, we propose a novel tree data structure called MedBlockTree and the corresponding consensus protocol, EnhancedPro, to address the scalability limitations in blockchain-based EMR systems, particularly on block throughput. By implementing the identity-based chameleon hash algorithm, MedBlockTree not only ensures patients' awareness of newly generated blocks but also expands the chain structure into a growing tree, which can help parallelly process multiple blocks in a consensus round. EnhancedPro maintains network consistency by enabling MedBlockTree workers to independently process block proposers' elections using recognized parameters, while implementing a Byzantine Fault Tolerance mechanism to mitigate potential network disruptions. The simulation results show that the block throughput (BPS) / overall processing time of MedBlockTree is significantly enhanced compared with blockchain-based systems throughout branch numbers, network latency, network size, and the collision rates. The findings highlight MedBlockTree's potential as a robust and scalable solution for secure medical data sharing, with the potential to drive the evolution of blockchain applications in healthcare informatics. 

However, as the number of branches increases, we observe a pattern of diminishing returns in BPS, suggesting a plateau in scalability benefits with further increases. Future work will focus on developing efficient branch management strategies to address this limitation.

\section*{Acknowledgment}
This work is supported by the China Scholarships Council and the Victoria University of Wellington Doctoral Scholarship. We extend our gratitude to Catalyst Cloud for their generous support through a student research grant. We also appreciate the valuable assistance and insights from our colleagues in the Wireless Network (WiNe) Research Group at Victoria University of Wellington.

\end{document}